\documentclass[11pt]{article}
\usepackage[pagewise]{lineno}
\newcommand{\comm}[1]{}

\usepackage{amssymb}
\usepackage{amsmath}\usepackage{graphicx} \usepackage[numbers]{natbib}\usepackage{epsfig}
\def\citet{\cite}

\def\xxxonly{\comm}

\usepackage{times}\hfuzz=10pt 

\newtheorem{theorem}{Theorem}
\newtheorem{lemma}{Lemma}
\newtheorem{corollary}{Corollary}
\newtheorem{definition}{Definition}
\newtheorem{remark}{Remark}
\newtheorem{example}{Example}

\def\e{\varepsilon}

\def\defi{\stackrel{{\scriptscriptstyle \Delta}}{=}}
\def\simone{\stackrel{{\scriptscriptstyle (1)}}{\simeq}}
\def\simtwo{\stackrel{{\scriptscriptstyle (2)}}{\simeq}}
\def\simp{\stackrel{{\scriptscriptstyle (p)}}{\simeq}}
\def\simmn{\stackrel{{\scriptscriptstyle (n)}}{\simeq}}
\def\simnmin{\stackrel{{\scriptscriptstyle (n-1)}}{\simeq}}

\def\d{\delta}
\def\o{\omega}
\def\O{\Omega}

\def\F{{\cal F}}
\def\w{\widehat}
\def\Ind{{\mathbb{I}}}

\def\Re{{\rm Re\,\!}}

\def\R{{\bf R}}

\def\F{{\cal Z}}

\def\L{L}
\def\F{{\Feta}}

\def\C{{\bf C}}

\def\oo{\bar}

\def\U{{\cal U}}

\def\L{{\cal L}}
\def\I{{\,\! \cal I}}

\def\F{{\cal F}}

\newcommand{\be}{\begin{equation}}
\newcommand{\ee}{\end{equation}}
\newcommand{\bd}{\begin{displaymath}}
\newcommand{\ed}{\end{displaymath}}
\newcommand{\ba}{\begin{array}{ll}}
\newcommand{\ea}{\end{array}}
\newcommand{\baa}{\begin{eqnarray}}
\newcommand{\eaa}{\end{eqnarray}}
\newcommand{\baaa}{\begin{eqnarray*}}
\newcommand{\eaaa}{\end{eqnarray*}}

\def\ZZ{{\bf Z}}

\def\oo{\bar}

\def\CC{{\cal C}}

\def\ew{\left(e^{i\o}\right)}
\def\ew{\left(i\o\right)}

\def\TT{{\cal T}}

\def\ee{\epsilon}

\title{
Spectrum degeneracy for functions on branching lines and impact on extrapolation and sampling
}
\author{
Nikolai Dokuchaev}
 
\begin{document}
 \vspace{-0.5cm}      \maketitle
\def\brea{}
\def\breakk{}
\def\break{}
\def\break{\nonumber\\ }\def\breakk{\nonumber\\&&}\def\brea{\nonumber\\ }
\let\thefootnote\relax\footnote{The author is with the School of Electrical Engineering, 
Computing and Mathematical Sciences, Curtin University, GPO Box U1987, Perth,
Western Australia, 6845 (email N.Dokuchaev@curtin.edu.au). }
\begin{abstract}
The paper studies functions defined on continuous branching lines connected into a system.
\index{The topology of the system is taken into account via the restriction that these processes coincides on certain
parts of the real axis.} 
A notion of spectrum degeneracy for these  functions is introduced. This degeneracy is based on the properties of the
Fourier transforms for processes representing  functions on the branches.\index{ that are deemed to be extended  onto the real axis. }  It is shown that processes with this spectrum degeneracy are everywhere dense in the set of processes equivalent
to functions on the branching lines. Some applications to extrapolation and sampling are considered.
\end{abstract}
\section{Introduction}
The paper considers functions defined on continuous branching lines connected to a system and spectrum degeneracy for them  in the pathwise deterministic setting, i.e.  without probabilistic assumptions on the ensemble.

It is known that there are some   opportunities for prediction and interpolation
of continuous time processes  with  certain degeneracy of
their spectrum. Let us list some results known in the pathwise setting.  The classical
sampling theorem  states that
 a band-limited continuous time function can be uniquely recovered without error  from  a  sampling sequence   taken with sufficient frequency.  Continuous time functions with periodic gaps in the spectrum
can be recovered from sparse samples;
see \cite{La2,OU08,OU}. Continuous time band-limited functions are analytic and can be recovered from the values on an arbitrarily small time interval.
In particular, band-limited functions can be predicted from their past values. Continuous time functions with the Fourier transform vanishing on an arbitrarily small interval $(-\O,\O)$ for some $\O>0$  are uniquely defined by their past values; there are
linear predictors that do not require to know the spectrum allowing to predict anticausal convolutions \citet{D08}.

It appears that many applications require to extend the existing theory on the processes defined on the domains with a non-trivial topological structure. Currently, the main efforts
are directed  toward signal processing on graphs in the discrete setting  based on the sampling on the vertices; see a review on
\citet{SN}.  The present  paper considers a different setting with domain represented by  continuous branching lines.
Since the branches are not homeomorphic to $\R$, it is not obvious how to introduce the notion   of spectrum degeneracy for these domains. In addition, it is required to take into account the topology of the branching. The paper suggests an approach that allows to use the Fourier transform or processes representing  functions on the branches that are deemed to be extended  onto the real axis.
The topology of the system is taken into account via the restriction that these processes coincides on certain
parts of the real axis.
It is shown that processes with the  spectrum degeneracy of the suggested kind are everywhere dense in the set of the underlying processes  (Theorem \ref{ThDense}). Some applications to extrapolation and sampling are considered (Theorem \ref{ThU} and Corollary \ref{corrS}). For example, it appears that a function defined on a tree allows an arbitrarily close approximation by a function
that is uniquely defined by its equidistant sample
sample taken on a semi-infinite half of a root (Example \ref{corrT}).

\section{Definitions}

For complex valued functions $x\in L_1(\R)$ or $x\in L_2(\R)$, we
denote by $\F x$ the function defined on $i\R$, where $i=\sqrt{-1}$, as the Fourier
transform of $x$; $$(\F x)(i\o)= \int_{-\infty}^{\infty}e^{-i\o
t}x(t)dt,\quad \o\in\R.$$ If $x\in L_2(\R)$, then $X(i\cdot)$ is defined
as an element of $L_2(\R)$ (meaning that  $X\in L_2(i\R)$). If $X(i\cdot)\in L_1(R)$ then
$x=\F^{-1}X\in C(\R)$ (i.e. it is a bounded and continuous function on $\R$).

Let $m>0$ be a fixed integer.

Let $\I$ be the set of all measurable sets $I\subset\R$ such that there exists $a\in\R$ such that  either $(-\infty,a)\subset I$  or $(a,\infty)\subset I$.

 Let $S$ be the set of all sets $\TT \subset \{1,...,m\}^2\times \I$ such that if $(d,k,I)\in\TT$ then
 $d\neq k$.

\begin{definition} \label{def1} \begin{enumerate}
\item
For a given $\TT\in S$, let $\L_{2,\TT}$ be the set of all
 ordered sets $\{ x_d\}_{d=1}^{m}\in [L_2(\R)]^m$ such that  $(d,k,I)\in\TT$ if and only if
 $x_d|_I=x_k|_I$ up to equivalency (i.e.  $x_d(t)=x_k(t)$ almost everywhere on $I$).
\item For a given $\TT\in S$, let $\CC_{\TT}$ be the set of all  $\{ x_d\}_{d=1}^{m}\in \L_{2,\TT}$ such that $x_d\in C(\R)$ for all $d$.
\item For a given $\TT\in S$, let $\w\CC_{\TT}$ be the set of all
 ordered sets $\{ x_d\}_{d=1}^{m}\in \CC_{\TT}$ such that  $X_d(i\cdot)\in L_1(\R)$ for all $d$, where
 $X_d=\F x_d$.
 \end{enumerate}
In all these cases, we say that $\{ x_d\}_{d=1}^{m}$ from Definition \ref{def1}  is  a branching process with the  structure set  $\TT$.
\end{definition}


\begin{definition}\label{defBranch} For $\TT\in S$ and $n\in\{1,...,m\}$, let us define a relation $\simmn$  between $d\in\{1,...,m\}$ and $k\in\{1,...,m\}$ recursively
as the following.
\begin{itemize}
\item[(a)]  $d\simone k$  if and only  if $(d,k,I)\in\TT$ for some $I\in \I$.
\item[(b)] For $n\in\{2,...,m\}$,  $d\simmn k$ if and only if \index{the following two conditions are satisfied:
\subitem$-$  for any $p\in\{1,...,n-1\}$,
it is not true that $d\simp k$,  and
\subitem$-$} there exists  $q\in\{1,...,m\}$ such that $d\simone q$ and $q \simnmin k$.
 \end{itemize}
We say that $d\simeq k$ if and only if  there exists $n\in\{1,...,m\}$ such that $d\simmn k$.
 \end{definition}

In particular, the  relation $\simeq$ is symmetric and  transitive.

\begin{remark} A branching process  from Definition \ref{def1}
can be associated with  a function on a branching line (i.e. an one dimensional manifold representing a branching
structure)  as the following: for each  $(d,k,I)\in\TT$,  the process $(x_d,x_k)$ represents
the  process $\{(x_d(t),x_k(t))_{t\in \R\setminus I}, x_d(t)|_{t\in I}\}$; this can be done since, by the definitions $x_d|_I=x_k|_I$.
\end{remark}

Up to the end of this paper, we  consider branching processes  $\{ x_d\}_{d=1}^{m}$ such as described in Definition \ref{def1}, presuming that they describe functions on the corresponding
branching lines.

\begin{example}
A branching process  (or functions on branching lines) defined above can be used to describe  forecasting models for processes evolving in time such that the branching represents
different possible
scenarios. For example, let $t=0$ be a critical point after which a process can  evolve according to two different evolution laws.
This can be modelled by a branching process $\{x_d(t)\}_{d=1,2}$ from Definition \ref{def1}  with $\TT=\{(1,2, (-\infty,0),(2,1, (-\infty,0)\}$,
i.e. with   $x_1(t)=x_2(t)$ for $t<0$. \end{example}


For $\d>0$ and an interval $J\subset \R$, let $V(J)$ be the set of all   $x\in L_2(\R)$ such that
$X\ew=0$ for $\o\in J$, where $X=\F x$.

\begin{definition}\label{defD}  Let $\d>0$  be given. We say that a branching process
$\{ x_d\}_{d=1}^{m}\in \L_{2,\TT}$ features branching spectrum  degeneracy with the parameter $\d$  if  there exists a set of
intervals $\{J_d(\d)\}_{d=1}^{m}$ such that
  $ x_d\in V( J_d(\d))$  for all $d$, where  $J_d(\d)\defi (\o_d-\d,\o_d+\d)$, and where $\o_d\in\R$.
  We denote by $ \U_{\TT,\d}$ the set  of  all branching processes from $\L_{2,\TT}$ with this feature.
  \end{definition}
  \begin{remark} If $J_d(\d)\cap J_k(\d)\neq \emptyset$, then  $x_k\equiv x_d$, in the notations of Definition \ref{defD}; in particular, this follows from Lemma \ref{lemmaU} below.
For the proof of  Theorem \ref{ThDense}  below, we would need to consider  a more interesting  case where  the
intervals $\{J_d(\d)\}_{d=1}^{m}$ are disjoint.
\end{remark}


 \begin{definition}\label{defConn}
 We say that a  structure set  $\TT\subset \{1,...,m\}^2\times \I$
 is connected if $k\simeq d$ for all $k,d\in\{1,...,m\}$.
 \end{definition}
 \section{The main results}
 It is known that set of all band-limited processes is everywhere dense in $L_2(\R)$;
 this allows to  approximate a general type process by a band-limited one.
The following theorem establishes a similar property for branching processes.
\begin{theorem}\label{ThDense}
\begin{enumerate}
\item
For any $\TT\in S$, for any branching process
$\{x_d\}_{d=1}^{m}\in \L_{2,\TT}$, and for any $\e>0$, there exists  a branching process
$\{\w x_d\}_{d=1}^{m}\in \cup_{\d>0} \U_{\TT,\d}$  such that
\baaa
\max_{d=1,...,m}\|x_d-\w x_d\|_{ L_2(\R)}\le \e.
\label{xdd}\eaaa
\item
For any branching process
$\{x_d\}_{d=1}^{m}\in\w\CC_{\TT}$  and any $\e>0$, there exists  a branching process
$\{\w x_d\}_{d=1}^{m}\in [\cup_{\d>0} \U_{\TT,\d}]\cap\w\CC_{\TT}$  such that \baaa
\max_{d=1,...,m}(\|x_d-\w x_d\|_{ L_2(\R)}+\|x_d-\w x_d\|_{ C(\R)})\le \e.
\label{xdd2}\eaaa
\end{enumerate}
\end{theorem}

\par
According to Theorem \ref{ThDense}, the set  of branching processes featuring spectrum degeneracy
is everywhere dense in wide classes of branching processes. This leads to the possibility of applications to functions of a general kind on branching lines.

\begin{theorem}\label{ThU}   Any branching process
$\{x_d\}_{d=1}^{m}\in \U_{\TT,\d}$ with  a connected structure set $\TT\in S$
is uniquely defined by the path $x_k|_I$, for any $\d>0$, any $k\in\{1,...,m\}$, and any $I\in \I$. \end{theorem}

In Theorem \ref{ThU}, for a branching process from $\{x_d\}_{d=1}^{m}\in \U_{\d,\TT}$, the processes  $x_d$ are defined uniquely in $L_2(\R)$. For  a branching process  from $\U_{\d,\TT}\cap \C_{\TT}$,  $x_d$ is  defined uniquely  in $C(\R)$.

\begin{remark} Theorem \ref{ThU} claims an uniqueness result but does not suggest an method of extrapolation from the set from $\I$.
Some linear predictors allowing  the required extrapolation can be found in \xxxonly{\cite{D17} and} \cite{D08}.
\end{remark}

The following corollary represents a modification for branching processes of the classical sampling theorem
(Nyquist-Shannon-Kotelnikov Theorem).
\begin{corollary}\label{corrS} Let  $\d>0$, $\O>0$, and $\tau\in (0,\pi/\O)$ be given, and let a branching process
$\{x_d\}_{d=1}^{m}\in \U_{\TT,\d}$ be
such that
$X_1\ew=0$ for $\o\in (-\O,\O)$, where $X_1=\F x_1$.
Then, for any $s\in\ZZ$,  the branching process
$\{x_d\}_{d=1}^{m}$
is uniquely  up to equivalency defined by the sampling sequence  $\{x_1(t_k)\}_{k\in\ZZ,\ k\le s}$, where $t_k=\tau k$.
\end{corollary}

It is interesting that the sampling rate required  in  Corollary \ref{corrS} is independent on the parameter
$\d$ characterizing the spectrum degeneracy of the branching process in Definition \ref{defD}.
It can be also noted that  the processes $x_k$ in this corollary are not necessarily band-limited if  $k>1$.

It can be noted that the classical Nyquist-Shannon-Kotelnikov Theorem states that
a band-limited function $x\in L_2(\R)$ is uniquely defined by the  sequence $\{x(t_k)\}_{k\in \ZZ}$, where
$X\ew=0$ for $\o\notin (-\O,\O)$,  $X=\F x$, $t_k=\tau k$; this theorem allows $\tau \le \pi/\O$.
There is  a version of this theorem  for oversampling sequences  with $\tau <\pi/\O$: for any $s\in\ZZ$,  this $x$  is uniquely defined by the  sequence $\{x(t_k)\}_{k\in \ZZ, k\le s}$ \citet{F91,V87}.
Corollary \ref{corrS} extends this version on the case of branching processes: for any $s\in\R$,  the  branching process
$\{x_k\}$  is uniquely defined by the  sequence $\{x_1(t_k)\}_{k\in \ZZ, k\le s}$.

\begin{example}\label{corrT} Consider a branching line $T$ with a topology  corresponding to a tree, with a semi infinite root branch $T_0=\{t:\ t\le 0\}$. Let $f:T\to\R$ be a
function. Then, for any $\e>0$, there exists $\tau>0$ and a function $f_\e:T\to \R$ and $\d>0$ such that the following holds:
\begin{enumerate}
\item
$\sup_{t\in T}|f(t)-f_\e(t)|\le\e$,
\item
 for any $s<0$,  an  equidistant  sequence $\{f_\e(t_k)\}_{k\in\ZZ,\ k<s}$ defines $f$ uniquely for $t_k=\tau k$.
\end{enumerate}
\end{example}
\section{Proofs}

{\em Proof of Theorem \ref{ThDense}}.  Let us suggest a procedure for the construction of $\w x$; this will be sufficient to prove the theorem. This procedure is given below.

\par
For $d,k=1,...,m$,  let $Y_{k,d}\defi X_{k}-X_{d}$ and $y_{k,d}=\F^{-1}Y_{k,d}$, where $ X_{k}=\F  x_{k}$.

Let a set $\{\o_{k}\}_{k=1,...,m}$ be such that all its elements are different. Consider  a system of intervals
 $\{J_{k}(\d)\}_{k=1,...,m}$  such that   $J_{k}(\d)=(\o_{k}-\d,\o_{k}+\d)$.
 We assume below that $\d>0$ is small enough  such that these intervals are disjoint.

Set
\baaa
\w X_1\ew\defi X_1\ew\Ind_{\{\o\notin \cup_{d=1}^{m} J_{d}(\d)\}} -
\sum_{d=2}^m Y_{d,1}\ew\Ind_{\{\o\in J_{d}(\d)\}}\eaaa
and
\baaa
\w X_d\defi \w X_1+ Y_{d,1}, \quad d=2,...,m.
\eaaa
\par
 Assume that $(d,k,I)\in\TT$. We have that
\baaa
\w X_k-\w X_d= \w X_1+Y_{k,1}-\w X_1+ Y_{d,1}\brea
=\w X_{k,1}-\w X_1-\w X_d+\w X_{1}=Y_{k,d},
\eaaa
i.e.
\baaa
\w X_k=\w X_d+Y_{k,d}.
\eaaa
Let $\w x_d=\F^{-1}\w X_d$, $d=1,...,m$.

Under the assumptions of statement (i) of the theorem, we have that  $x_{k}|_I=x_{d}|_I$ up to equivalency. It follows that $y_{k,d}|_I=0$ up to equivalency, i.e.  $\w x_d|_I=\w x_k|_I$ up to equivalency.
Similarly, under the assumptions of statement (ii) of the theorem, we have that  $x_{k}|_I=x_{d}|_I$. It follows that $y_{k,d}|_I=0$, i.e.  $\w x_d|_I=\w x_k|_I$.
Since this holds for all   $(d,k,I)\in\TT$, it follows that $\{\w x_d\}_{d=1}^m$ is a branching process with the same structure set $\TT$ as the underlying branching process $\{ x_d\}_{d=1}^m$.

Let us show  that the branching process
$\{\w x_d\}_{d=1}^{m}$ features the required  spectrum degeneracy.
Since the sets $J_d(\d)$ are mutually disjoint,
 it follows immediately from the definition for $\w X_1$ that   $\w X_1\ew =0$ for $\o\in J_1(\d)$.
 Further, by the definition for $\w X_d$ for $d>1$, we have that
\baaa
\w X_d\ew=\w X_{1}\ew+Y_{d,1}\ew\brea= X_1\ew\Ind_{\{\o\notin \cup_{d=1}^{m} J_{d}(\d)\}} -
\sum_{d=2}^m Y_{d,1}\ew\Ind_{\{\o\in J_{d}(\d)\}}\brea+Y_{d,1}\ew, \eaaa
Since the sets $J_d(\d)$ are mutually disjoint,
 it follows again that   $\w X_d\ew =0$ for
$\o\in J_d(\d)$  for all $d$. Hence $\w x_d\defi\F^{-1}\w X_d\in V(J_d(\d))$
for
all $d=1,...,m$.  It follows that the branching  process
$\{\w x_d\}_{d=1}^{m}$ belongs to $\U_{\TT,\d}$, i.e. features the required spectrum degeneracy.

Furthermore, for all $d$,  we have that, under the assumptions of statement (i),
$\|X_d(i\cdot)-\w X_d(i\cdot)\|_{L_2(\R)}\to 0$ as $\d\to 0$.
In addition,    we have that, under the assumptions of statement (ii),
 \baaa
 \|X_d(i\cdot)-\w X_d(i\cdot)\|_{L_2(\R)}+\|X_d(i\cdot)-\w X_d(i\cdot)\|_{L_1(\R)} \to 0\eaaa
 as $\d\to 0$. Under the assumptions of statement (i), it follows that $\|\w x_d-x_d\|_{ L_2(\R)}\to 0$. Under the assumptions of statement (ii), it follows that  $\|\w x_d-x_d\|_{ L_2(\R)}+\|\w x_d-x_d\|_{ L_2(\R)}\to 0$  as $\d\to 0$.
This completes the proof of Theorem \ref{ThDense}. $\Box$

Let us prove Theorem \ref{ThU}. For the case where $m=1$ (this case is not excluded),
the statement of Theorem \ref{ThU}  is known and can be reformulated as  the following lemma; for completeness, we provide its proof.
\begin{lemma}\label{lemmaU} Let $\U_1$ be the set of all  $x\in  L_2(\R)$ such that there exists an interval
$(a,b)\subset \R$ such that  $X\ew =0$ for $\o\in (a,b)$, $X=\F x$, $a<b$.
Then any $x\in \U_1$ is uniquely defined by its path $x|_I$ for any $I\in\I$. \end{lemma}

{\em Proof  of Lemma \ref{lemmaU}}. Without loss of generality, we assume that
 $(-\infty,0)\subset I$. Let
$\C^+\defi\{z\in\C:\ \Re z>  0\}$, and let  $H^2$ be the Hardy space of holomorphic on $\C^+$ functions
$h(p)$ with finite norm
$\|h\|_{H^2}=\sup_{s>0}\|h(s+i\o)\|_{L_2(\R)}$; see, e.g. \cite{Du}, Chapter 11.
 It suffices to
prove that if $x\in L_2(\R)$ is such that $X\ew =0$ for $\o\in (a,b)$, $X=\F x$, $a<b$, and  $x(t)=0$ for
$t\le 0$,  then  $x(t)=0$ for $t>0$. These properties imply that $X\in H^2$, and, at the same time, \baaa \int_{-\infty}^{+\infty}(1+\o^2)^{-1}|\log|X\ew||dx=+\infty.\eaaa
 Hence, by the property of the Hardy space,
  $X\equiv 0$; see, e.g.  Theorem 11.6 in \cite{Du}, p. 193.
This completes the proof of Lemma \ref{lemmaU}. $\Box$

{\em Proof  of Theorem \ref{ThU}}. Let $I\in \I$. By Lemma \ref{lemmaU},  $x_1$ is uniquely
defined by $x_1|_I$. Further, let $k\simone 1$, then $x_k$ is uniquely
defined by $x_k|_I=x_1|_I$. Similarly, if $k\simtwo 1$, $k\simone q$, and $q\simone 1$,
then $x_k$ is uniquely
defined by $x_q|_{\oo I}$ for some $\oo I\in\I$, and $x_q$  is uniquely defined by $x_1|_I$.
It follows that $x_k$  is uniquely defined by $x_1|_I$ as well.
Extending this approach on all $k$, we similarly obtain  the statement of the theorem.
This completes the proof of Theorem \ref{ThU}. $\Box$

{\em Proof  of Corollary \ref{corrS}}. It follows from  the results \citet{F91,V87}
that $x_1$ is  uniquely  defined by $\{x_1(t_k)\}_{k\le s}$. Then the statement of Corollary \ref{corrS}
follows from Theorem \ref{ThU}. $\Box$

\section{Conclusions and future research}\label{secCon}
The present paper is focused on the frequency analysis  for   functions defined
on continuous branching lines, i.e. lines connected to a system.
The paper suggests an approach that allows to take into account
the topology of the brancing line via modelling it  as a system of standard processes  defined on the real axis and
coinciding on certain intervals  with well-defined Fourier transforms (Definition \ref{def1}). This approach can be applied to a variety of branching lines. In some cases, it may require a straightforward  modification of the underlying system.
 In particular, to apply our approach to a system with   compact  branching lines,
it suffices to extend the domain of these functions. For example,  one can  extend
edges of the branching line beyond  their vertices. This would mean transformation of finite edges into semi-infinite ones. Alternatively,  one can supplement the branching line by new dummy semi-infinite edges originated from the vertices of order one.

We leave for the future research classification and   description of branching lines allowing to represent functions defined on them via
branching processes from Definition \ref{def1}.

\xxxonly{A result similar to Theorem \ref{ThU} for a class of processes similar to processes from
$\U_{\TT,\d}$ but with spectrum for $x_d$ vanishing at single points with sufficiently high vanishing rate
such as described in \cite{D17}.}

 \end{document}